\documentclass[epj]{svjour}
\usepackage{graphics, caption}
\usepackage{bm}
\usepackage{amsmath}
\usepackage{amsfonts}
\usepackage{cancel}
\usepackage{xcolor}
\usepackage{xspace}
\usepackage{slashed}
\usepackage{hyperref}

\def\eg{\emph{e.g.}\xspace}
\def\ie{\emph{i.e.}\xspace}
\def\jets{\ensuremath{\mathrm{jets}}}
\def\hgg{$\mathrm{H}\to\gamma\gamma$\xspace}
\def\gj{$\gamma + \jets$\xspace}
\def\CR{\ensuremath{\mathrm{CR}}\xspace}
\def\SR{\ensuremath{\mathrm{SR}}\xspace}
\newcommand\pt[1]{\ensuremath{{p_\mathrm{T}}_{#1}\xspace}}
\newcommand\id[1]{\ensuremath{\mathrm{ID}_{#1}}\xspace} 

\def\gam{\ensuremath{\gamma}\xspace}
\def\misgam{\ensuremath{\slashed{\gamma}}\xspace}
\begin{document}
\title{Data driven background estimation in HEP using Generative Adversarial Networks}
\author{Victor Lohezic\inst{1}, Mehmet Ozgur Sahin\inst{1}, Fabrice Couderc\inst{1} \and Julie Malcles\inst{1} 
%
}                     
%
%
\institute{IRFU, CEA, Université Paris-Saclay, F-91191 Gif-sur-Yvette, France}
\date{Received: 09.12.2022}
%
\abstract{
Data-driven methods are widely used to overcome shortcomings of Monte Carlo simulations (lack of statistics, mismodeling of processes, etc.) in experimental high energy physics. A precise description of background processes is crucial to reach the optimal sensitivity for a measurement. However, the selection of the control region used to describe the background process in a region of interest biases the distribution of some physics observables, rendering the use of such observables impossible in a physics analysis. Rather than discarding these events and/or observables, we propose a novel method to generate physics objects compatible with the region of interest and properly describing the correlations with the rest of the event properties. We use a generative adversarial network (GAN) for this task, as GANs are among the best generator models for various applications. We illustrate the method by generating a new misidentified photon for the $\gamma + \mathrm{jets}$ background of the $\mathrm{H}\to\gamma\gamma$ analysis at the CERN LHC, and demonstrate that this GAN generator is able to produce a coherent object correlated with the different properties of the rest of the event.
}
\authorrunning{V. Lohezic, M. O. Sahin, F. Couderc \and J. Malcles} 
\titlerunning{Data driven background estimation in HEP using GAN}
\maketitle

\section{Introduction}
\label{sec:intro}

In high energy physics (HEP), characterizing a signal hypothesis requires distinguishing its signature from a large number of background processes with similar final states. 
Observables of physics objects are used to construct classification algorithms that can discriminate the signal signatures from the background processes. 
An accurate description of these final states and their observables with the detailed modelling of the detector responses is crucial for the sensitivity of the analysis. \\
Monte Carlo techniques are widely used in HEP experiments to simulate a process from a physics model of interest (denoted as signal) and other Standard Model (SM) processes (denoted as background). 
Software libraries such as GEANT4~\cite{GEANT4_2003,GEANT4_2016} are used to model detailed descriptions of the modern colossal particle detectors, and thereby an accurate simulation of the detector responses to the aforementioned processes.
Due to the very intricate nature of these detectors, the large amount of data delivered by the colliders and the rarity of the signal, a substantial computational infrastructure in the form of grid processing power and a significantly large storage-disc volume is needed.
This requirement usually constrains the simulation sample production to have limited statistics in the tails of discriminating observables. 
 
Furthermore,
inaccuracies in the underlying physics model and/or in the description of detector responses are limiting factors to the use of MC simulations for background description. 
Data driven techniques are used to overcome these challenges. 

\begin{figure*}[!h]
\begin{center}

\resizebox{0.95\textwidth}{!}{%
  \includegraphics{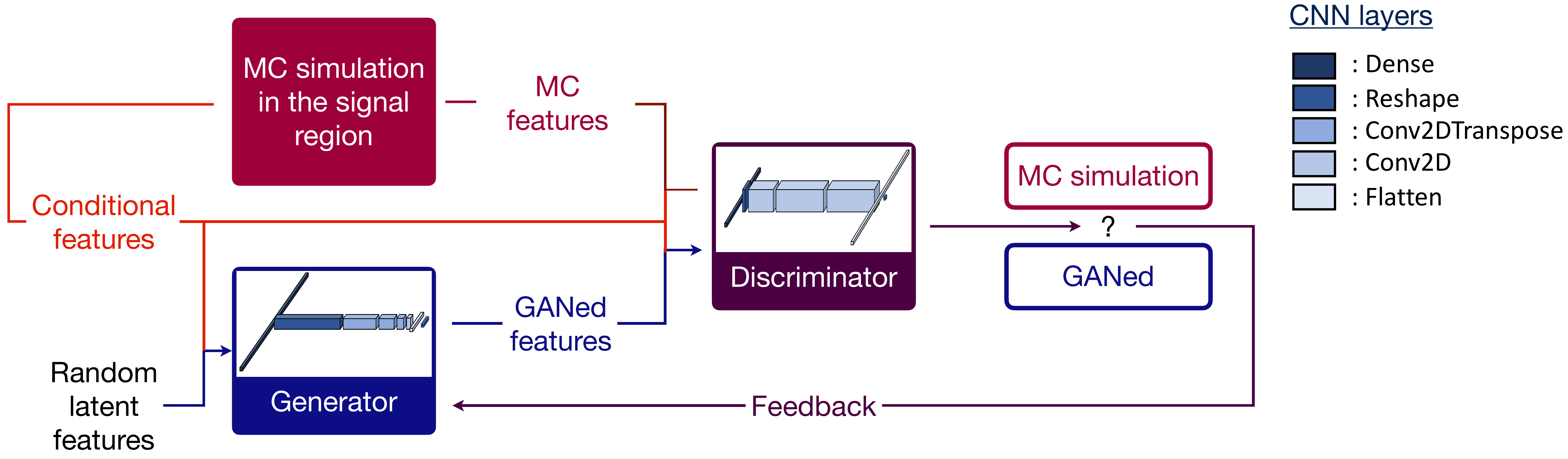}
}
\caption{Layout of the proposed conditional GAN architecture training. Conditional features and random latent features are given as inputs to the generator model. The generator model creates a new object with properties of a misidentified object of the \SR. The discriminator model is expected to classify between MC-simulated events and GANed events.}

\label{fig:GAN}    
\end{center}
\end{figure*}
ML classification algorithms are widely used to obtain the optimal separation between signal and background processes. 
They can extract the higher order relations between observables to provide a better classification performance~\cite{Radovic} with respect to techniques treating observables sequentially.
Therefore, the training samples should provide a good description of signal and background observables and their correlations.
Many of the leading background processes have signatures mimicking the signal due to one or more misidentified particles in their final states.
For instance, the Higgs boson signature with two isolated photons in the final state will have to be distinguished from other SM processes with single photon and multiple jets in the final state where one of the jets is misidentified as a photon. 
A similar example can be given for the signal signatures where two b jets are expected. In this case, processes with a single b jet can have a second light flavor jet misclassified as b jet, thus populating the signal region.\\
Modelling misidentification of the objects is a challenging task as it might be subject to systematic effects creating discrepancies at the extremes of distributions. 
Various data driven techniques are used to mitigate possible impacts of mismodelling.
For instance a data driven technique that requires two more additional sub-sample spaces, known as ABCD technique, is widely used in HEP, \eg in Ref.~\cite{PhysRevD.44.29}.
Even though this technique is quite capable of estimating the yields of the different processes, the shapes or the correlations of the observables in the high purity signal region cannot be retained. 
Other techniques are also specifically tailored for particular signatures.
However, their generalization cannot be assured, while they may still suffer from the aforementioned shortcomings.

In this paper, we propose a novel data-driven technique using a conditional generative adversarial network (GAN) to model backgrounds with misidentified particles as illustrated in Fig.~\ref{fig:GAN}.
We generate new observables for a particle that fails the identification criteria, so that it can mimic a misidentified object whose observables retain correlations with other event observables. This way, a complete event is generated for modeling the process in the region of interest by making use of discarded data events, which differentiates the proposed technique from the similar studies documented in Ref.~\cite{Lin:2019aa,Chisholm:101007,PhysRevD.106.055006}. Furthermore, we use all features that may have correlation with the generated observables as conditional inputs to the GAN rather than considering only a single feature (such as reconstructed mass of a particle, cf.~\cite{Chisholm:101007}), enabling the network to model linear and non-linear correlations within the generated object features and with the conditional features.

We will demonstrate the technique on a CMS open simulation sample described in Ref.~\cite{hgg_opendata}  consisting of the \gj process in the context of the CMS \hgg analysis as presented in Ref.~\cite{hgg_run1,hgg_run2}.

The paper is organized as follow: a brief introduction to GANs and related work in HEP will be given in Section~\ref{sec:gan}. The proposed technique will be laid out in the same section. The performance of this method will be assessed in Section~\ref{sec:appres}. 

\section{Generative Adversarial Networks}
\label{sec:gan}
A GAN consists of two neural networks competing in a zero sum game \cite{GAN}. 
In its basic form, the generator network takes a latent space of randomly distributed features.
The discriminator network compares the generator output with the input sample to distinguish between the generated and the original samples. 
The two networks are trained consecutively at each iteration (epoch) by minimizing loss metrics (one per network). A loss metric describes the difference between the objective and the value returned by the network. 
The generator produces samples more and more similar to the input sample with the feedback it received from the discriminator network.
Ideally, after a certain number of epochs the discriminator and generator networks should reach an optimal state where no further improvement can be obtained with the given dataset and number of parameters of the networks. 
Practically these networks may not reach a state of Nash equilibrium where further training does not change the performance of the two networks~\cite{pmlr-v119-farnia20a}. 

An imbalanced classification loss between two networks may cause training of a GAN to end before the optimum performance is achieved, particularly when the feedback loss from one network collapses into a constant value. To prevent this, it is imperative to have balanced losses, with the discriminator providing feedback to the other network at each training iteration. Because the losses of the two networks are balanced, an independent figure-of-merit is required to assess the performance of the generator.

In this paper, we propose a performance score based on the negative log-likelihood of the underlying distributions of the generated and real samples. 
We demonstrate that this score suffers from large fluctuations, intrinsic to the nature of GANs, and present a way to stabilize it. This allows one to pick the best performing model without introducing a significant computational overhead.

GANs are widely explored to provide a solution for the high computational requirements of the simulated sample generation at collider experiments (\eg in Ref.~\cite{ArjMartinez_2020,Lin_2019,CaloGAN}). 
They are also proposed to improve and generalize data-driven background techniques such as the ML based ABCD method~\cite{ABCDisco}.
In this work, we propose an alternative usage of GANs for a data driven technique. 

\subsection{Methodology}
After collecting data and constructing the observables, a typical analysis flow in HEP experiments starts with identifying the physics objects. 
Multivariate analysis techniques are widely used to provide an identification (\id{}) score. For instance, a photon \id{} score is developed to discriminate real prompt photons (originating from the primary vertex) from jets reconstructed as photon (named misidentified hereafter) in the \hgg analysis. Object candidates passing a certain threshold on the \id{} score are identified as physics objects (\eg signal photons).
Selection criteria are applied to choose events that have similar final states to the signal process of interest, creating a signal region named \SR. 
Various background processes with similar signatures may pass these selection criteria. 
Using the features of the selected objects, a multivariate technique can further be used to increase the signal purity in the \SR. 

Due to misidentified objects other background processes may also populate the \SR. 
For instance, any background process with a single b-tagged jet may appear in the search region of  $\mathrm{H\to b\overline{b}}$ signal process due to misidentification of a light flavor jet as b jet. 
In a wide variety of physics analyses such background processes constitute a large part of the overall background processes. 

In this paper, we propose a GAN-based approach to simulate such background processes using a data-driven technique. The method relies on selecting a control region (CR) using data events where one object fails the object selection criteria (\eg the photon \id{} criteria in \hgg analysis). In the \CR, the object failing the selection criteria will be replaced by a GAN-generated one.

This way, the \CR, where the object failing the \id{} criteria is replaced with a generated one, can simulate events from the same physics process in the \SR, \ie with a misidentified object. 
Furthermore, the correlations between the observables are retained. 
Another advantage is the gain in statistics as the size of the data sample with events failing the selection criteria is typically larger than the MC samples after the \SR selection, \eg jets passing stringent photon \id{} criteria are extremely rare.

\subsection{Model architecture}
The GAN model used in the present paper is an extension of the deep convolutional GAN (DC-GAN) architecture from Ref.~\cite{DCGAN}. 
A conditional generator network is adapted to retain the correlations of the input features with themselves and with other event-based observables. 
To achieve this, event-based observables, for which the correlations are desired, are used as input to both the generator and discriminator networks as shown in Fig.~\ref{fig:GAN}. These observables are called conditional features. A more detailed description of the networks architecture is given in section~\ref{subsec:trainval} for a specific application.

\subsection{Training}
Training and validation samples are taken from a MC simulation of the process of interest in the \SR. 
For each event $i$, a set $\bm{x}_i$ of $n_\mathrm{feat}$ features is defined. From this vector, 
a subset of conditional features $\bm{x}_{i,\mathrm{cond}}$ of size $n_\mathrm{cond}$ is selected and concatenated to a vector of random latent features sampled from a normal distribution $\bm{z}_i$ of size $n_\mathrm{rand}$ to form an input vector to the generator model $g : \mathbb{R}^{n_\mathrm{cond} + n_\mathrm{rand}} \to \mathbb{R}^{n_\mathrm{feat}}$. 
The goal of the generator model is to produce a set of observables of size $n_\mathrm{out}\equiv n_\mathrm{feat} - n_\mathrm{cond}$ describing a misidentified object in the \SR (\eg transverse momentum \pt{}, pseudorapidity $\eta$, ...). 
These generated observables are in turn concatenated to the set of conditional observables forming an output vector $g(\bm{x}_{i,\mathrm{cond}}, \bm{z}_i)$. 
The output vectors are used and compared to the original observables $\bm{x}_i$ to train the discriminator model $d : \mathbb{R}^{n_\mathrm{cond} + n_\mathrm{out}} \to [0, 1]$.
To distinguish between the generated and original observables, the events are coupled to a discriminator label $l_i$ with the choice of 0 for the generated observables $(\bm{y}_i = g(\bm{x}_{i,\mathrm{cond}}, \bm{z}_i),\ l_i = 0)$ and 1 for the original ones $(\bm{y}_i = \bm{x}_i,\ l_i = 1)$.
An additional noise is added to the training labels as it has been shown to increase the chance of convergence for GAN~\cite{Smoothlabel}. 
The noise value is generated uniformly between 0 and a maximum $\epsilon$ meaning $l_i \in [0, 0+\epsilon]\cup[1-\epsilon, 1]$. 
Finally, the objective for the discriminator model is to return a value as close as possible to the input label (i.e. $d(\bm{y}_i) = l_i$).

The weights of events are updated after minimizing the loss for a given batch of events. The following loss functions $L_d$ and $L_g$ are respectively defined for the discriminator and generator networks: 
\begin{equation}
\begin{aligned}
    & L_d =  \frac{1}{N_\mathrm{batch}}\sum_{i=1}^{N_\mathrm{batch}}l_i\ln(d(\bm{y}_i)) + (1-l_i)\ln(1-d(\bm{y}_i)) \\
    & L_g = \frac{1}{N_\mathrm{batch}}\sum_{i=1}^{N_\mathrm{batch}}\ln d(g(\bm{x}_{i,\mathrm{cond}}, \bm{z}_i))
\label{Eq:loss}
\end{aligned}
\end{equation}
where $N_\mathrm{batch}$ is the batch size. 
Three optimizer algorithms are tested: the stochastic gradient descend algorithm \\(SGD)~\cite{10.2307/2236690}, the Adam optimizer~\cite{Adam} and the LAMB optimizer~\cite{LAMB}. 
The Adam optimizer is an update to the SGD algorithm as it relies on the first and second order moments of the gradients to create an automatic and parameter-wise scaling of the learning rate. 
The LAMB optimizer is an adaptation of the Adam optimizer where the normalization and scaling are done layer-wise, allowing training to be performed in larger batch sizes, thus decreasing the training time. 

\subsection{Performance metrics}
As the generator and discriminator models are trained adversarially, their loss functions do not reflect the absolute performance of each neural network.
A proper performance score is needed to evaluate the generator model after each training step.
One of the primary objectives of this study is to make the generator model learn the correlations between the observables of the event. 
Therefore, we chose a negative log-likelihood based metric to compare original and generated samples.

From an input sample $X=\{\bm{x}_1, ..., \bm{x}_N\}$ with $N$ events and a random latent space $Z$, an output sample $Y=\{\bm{y}_1, ..., \bm{y}_N\}=g(X, Z)$ is generated with the GAN. 
Events are distributed in bins numbered from 1 to $N_b$ which span a multi-dimensional space chosen to reflect the physics one wants to reproduce (in our case  \pt{}, $\eta$, \id{} score of the misidentified photon in the \hgg analysis, ${p_T}_{\gamma\gamma}/m_{\gamma\gamma}$).
We defined a log-likelihood performance metric as:
\begin{equation}
    -2\ln{\Lambda(X|Y)} = -2 \sum_{k}^{N_b} m_{k} \ln p_{Y}(k)
\label{Eq:NLL}
\end{equation}
where $p_{Y}(k) = n_{k}/N$ represents the probability of an event to fall in bin $k$ estimated from sample $Y$ and $m_{k}$($n_{k}$) is the number of $\bm{x}_i$($\bm{y}_i$) in bin number $k$. For each dimension in the log-likelihood, we transform the distribution of the corresponding variable to be uniform so we can use 10 bins by dimension which is enough to capture the variable shape while retaining a sufficient amount of data per bin.

This metric is computed on the training sample and on an independent validation sample after each epoch, to check for any overtraining effect.
The optimal state of the GAN is then chosen as the set of weights giving the lowest $-2\ln{\Lambda}$ value on the validation sample. 
\begin{figure}[t]

\resizebox{0.5\textwidth}{!}{ 
  \includegraphics{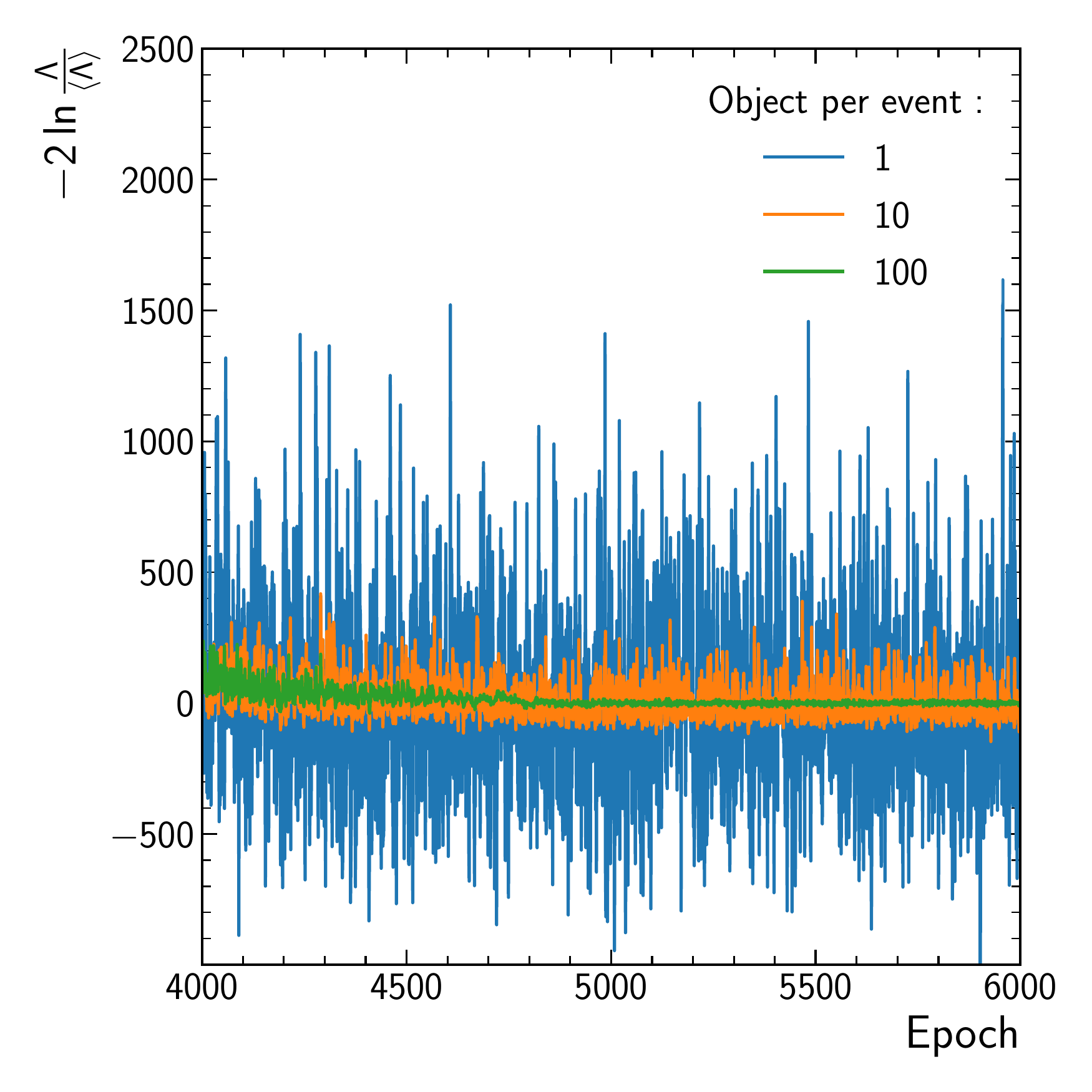}}

\caption{Comparison of the $\mathrm{-2\ln{\Lambda}}$ metric evolution during training for different numbers of objects generated per event using different random latent spaces. Each metric is shifted by its average over the last thousand epochs ($\mathrm{\left \langle \Lambda \right \rangle = \prod_{i=5000}^{6000} \Lambda_i}^{0.001}$, where $i$ refers to the epoch number) to better illustrate the stabilization.}

\label{fig:NgenPerEvent}       

\end{figure}
\subsection{Optimization strategy}

The balance needed between the performance of the two networks makes them prone to collapse towards suboptimal states, which produces a poor description of the event observables. 
Even when the GAN converges, large fluctuations are usually observed in the performance metrics. 
These fluctuations makes the optimization of the network challenging. 
An averaging method was developed to overcome this limitation and better assess the performance of the GAN. 
In the proposed GAN architecture, the features of an event are given as input to the model in addition to the the random latent space features. 
The GAN generator produces different objects for a given event (i.e. $g(\bm{x}_{i}, \bm{z}_i) \neq g(\bm{x}_{i}, \bm{z'}_i)$). 
Therefore, by generating multiple objects with different random features for each event, the performance metric of the GAN output is effectively averaged over the random latent space, giving a more accurate estimator $p_Y$ in Eq.~\ref{Eq:NLL}. 
This effect is demonstrated in Fig.~\ref{fig:NgenPerEvent} by using three different configurations where we generate 1, 10 and 100 objects. Fluctuations are stabilizing when more objects are generated per event. 
As the evaluation time increases with the number of generated objects, a compromise of 100 objects per event was chosen in this work.

With this reliable performance metric, different trained models can be compared when varying the hyperparameters used for each training. 
Several hyperparameters were tuned for this study:
\begin{itemize}
    \item networks architecture ($g$ and $d$)
    \item number of training events ($N$)
    \item gradient descent optimizer (SGD, Adam, LAMB)
    \item learning rate
    \item batch size ($N_\mathrm{batch}$)
    \item noise on training labels ($\epsilon$) of the discriminator
    \item set of conditional features ($\bm{x}_{i,\mathrm{cond}}$)
    \item random latent space dimension ($n_\mathrm{rand}$).
\end{itemize}
A description of this optimization and of the final set of hyperparameters for a concrete application are given in Section~\ref{subsec:trainval}.

In addition to this stabilization method and the tuning of hyperparameters, the preprocessing of the input features was studied. 
The goal of the preprocessing step is to transform the original input vectors into a representation more suitable for the training of neural networks. This transformation needs to be bijective so that a transformed vector can be processed back to its original values in a unique way. It can compensate rapidly falling or non smooth distributions of the physics observables --- those being harder to learn for a network. 
Multiple preprocessing methods are tested from the Scikit-learn module \cite{scikit-learn} and best performance is obtained with the quantile transformation to a uniform output.

\begin{figure}[ht]
\resizebox{0.5\textwidth}{!}{%
  \includegraphics{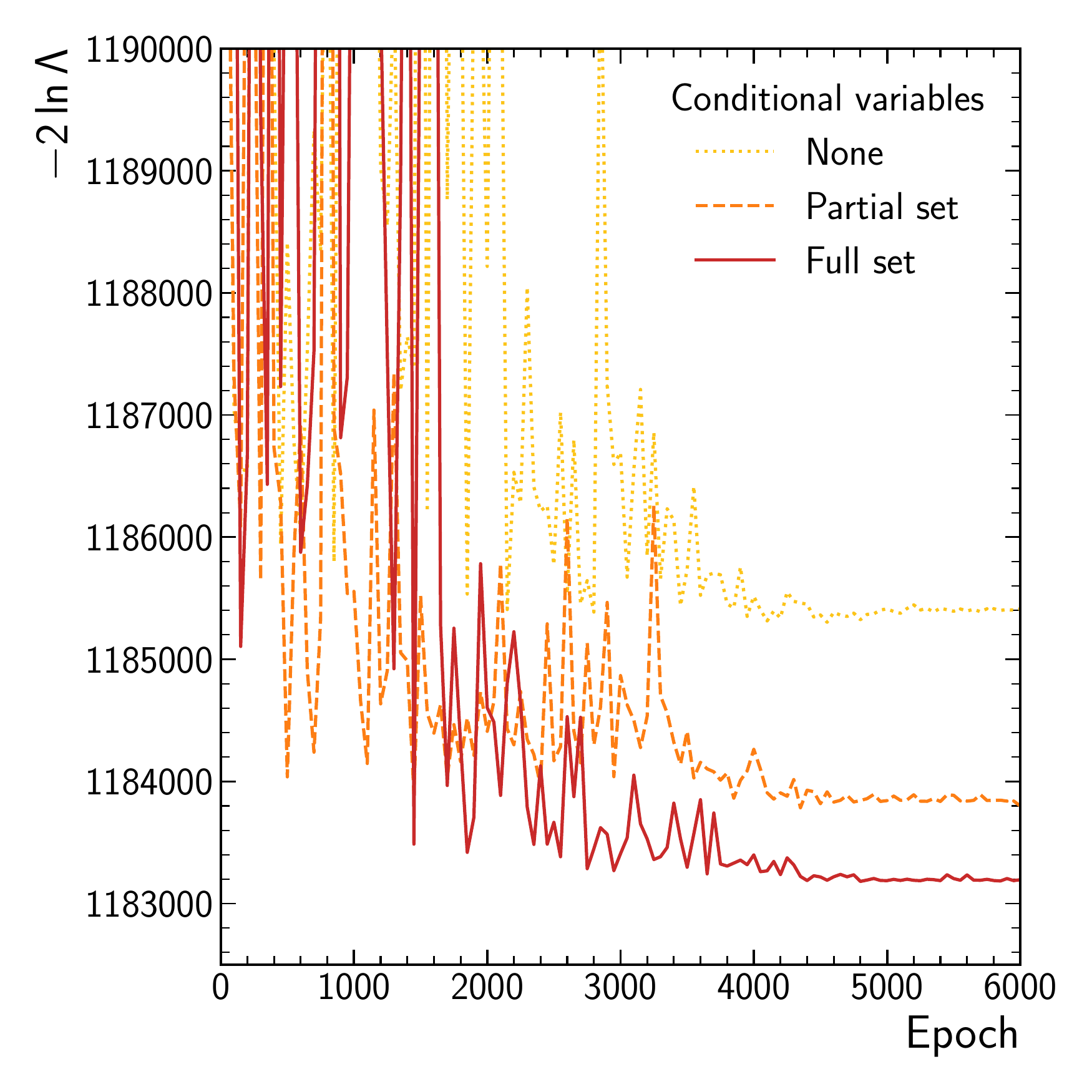}
}
\caption{Comparison of the $-2\ln{\Lambda}$ figure of merit evolution during the training of the different strategies tested for the addition of conditional features to the random latent ones.}
\label{fig:LossCond}       
\end{figure}

\section{Application and results}
\label{sec:appres}

Processes with a photon and multiple jets (\gj) in the final state are major backgrounds in the analysis of the Higgs-boson decay into diphoton \hgg.

In this analysis, photon candidates in the final state are selected by imposing photon \id{} criteria (\eg a threshold on a photon \id{} score). This defines the \SR in this work.
A jet passing the photon \id{} selection criteria is thus misidentified as a photon and hence contributes to the background in the \SR. 
The methodology described in section~\ref{sec:gan} can be used to estimate this background as in the following. 
Most data events with one jet failing photon selection criteria and another object satisfying this criteria are \gj or multi-jets events and constitute a control region with similar physics properties as the \SR, especially in terms of number of jets, jet identification, jet kinematics, jet flavours, etc. 
The GAN is trained with a \gj simulation to generate the observables of a misidentified photon (originating from a jet) in the \SR. 

Once trained, the GAN generator is used to replace the jet failing the photon \id{} criteria with a GANed misidentified photon, thus transforming an event residing in the \CR into an event located in the \SR. As the \CR with given criteria is exclusively composed of \gj events, the proposed technique allows evaluation of the \gj background in the \SR with a data-driven method. 
This is particularly handy when the analysis focuses on {\SR}s with additional selection criteria, \eg on the jet selection. Indeed in these cases the \gj simulation suffers from both a poor modeling of the physics and a lack of statistics.

To demonstrate this, a \gj simulated dataset from CMS open data~\cite{hgg_opendata} containing around 2.7 million events was used.

\subsection{Training and optimization}
\label{subsec:trainval}
Reconstructed momentum, position and the identification score are the observables of a misidentified object ( \pt{\misgam}, $\eta_\misgam$, $\phi_\misgam$, \id{\misgam}). 
The generator model creates these observables to replace the object failing the selection criteria in the control region with a GANed object in the \SR. 
Three different training strategies with respect to the feature sets are considered and compared.

The first strategy follows the vanilla GAN application as described in Section~\ref{sec:gan}, \ie the latent space is purely composed of random variables.
In the second strategy (partial set), we consider as conditional features the observables of the misidentified object. These conditional features together with the random latent ones are used as input to the GAN.
The last strategy (full set) takes as input random latent features and an extended set of conditional features: the observables of the misidentified object together with additional event observables like \id{\gam}, \pt{\gam}, $\eta_\gam$, $\phi_\gam$, $N_\jets$, $N_{\mathrm{vtx}}$.
We performed three trainings corresponding to these three strategies to test the impact of the conditional features.
As shown in Fig.~\ref{fig:LossCond}, the best performing strategy is to use the full set of conditional variables as the training loss reaches the lowest value of $-2\ln{\Lambda}$.
\begin{figure}[t]

\resizebox{0.49\textwidth}{!}{%
  \includegraphics{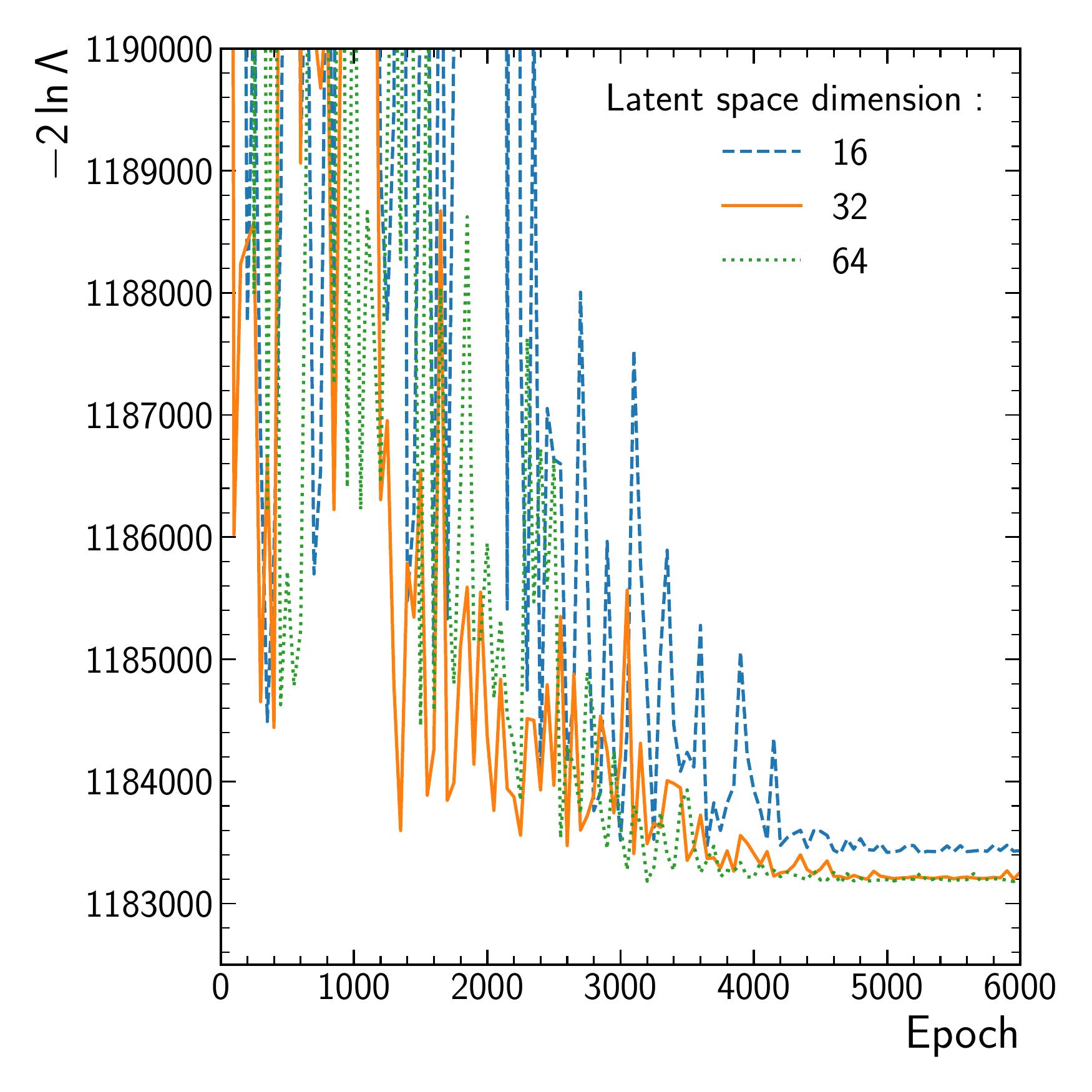}
}

\caption{Example of hyperparameter optimization where different configurations are tested. Lines with different colors correspond to different numbers of dimensions of the random latent spaces given as input to the generator model.}
\label{fig:LossLatent}       
\end{figure}

We further investigate the impact of different parameters of the models in the GAN. For instance, we find that increasing the dimension of the latent space to more than 32 does not provide additional performance improvement as demonstrated in Fig.~\ref{fig:LossLatent}.
Overall, the optimized parameters are found to be: 
\begin{itemize}
    \item random latent space dimension : 32
    \item number of training events : 100,000 events
    \item gradient descent optimizer : LAMB optimizer
    \item learning rate : Cosine decay as described in \cite{cosine_decay} starting at 0.001 and reaching 0 after 5,000 training epochs
    \item batch size : 1024
    \item noise on training labels : 0.15
    \item model architecture :
    \begin{description}
        \item[\textbf{generator} :] a dense input layer of 1024 nodes, three 2D deconvolution layers with 32/16/8 filters of size 4x4/2x2/2x2 respectively, one 2D convolution with 4 filters of size 3x3 and a dense layer with 4 outputs with hyperbolic tangent activation function
        \item[\textbf{discriminator} :] a dense input layer of 256 nodes, three 2D convolution layers with 32/64/64 filters of size 2x2/2x2/4x4 respectively and with a LeakyReLU activation function~\cite{leakyrelu}, a dense layer with 1 output with sigmoid activation function. A dropout~\cite{10.5555/2627435.2670313} of 20~\% is also implemented before the last layer of the discriminator.
    \end{description}
\end{itemize}

\subsection{Results}
\label{subsec:results}
\begin{figure}[t]

\resizebox{0.49\textwidth}{!}{%
  \includegraphics{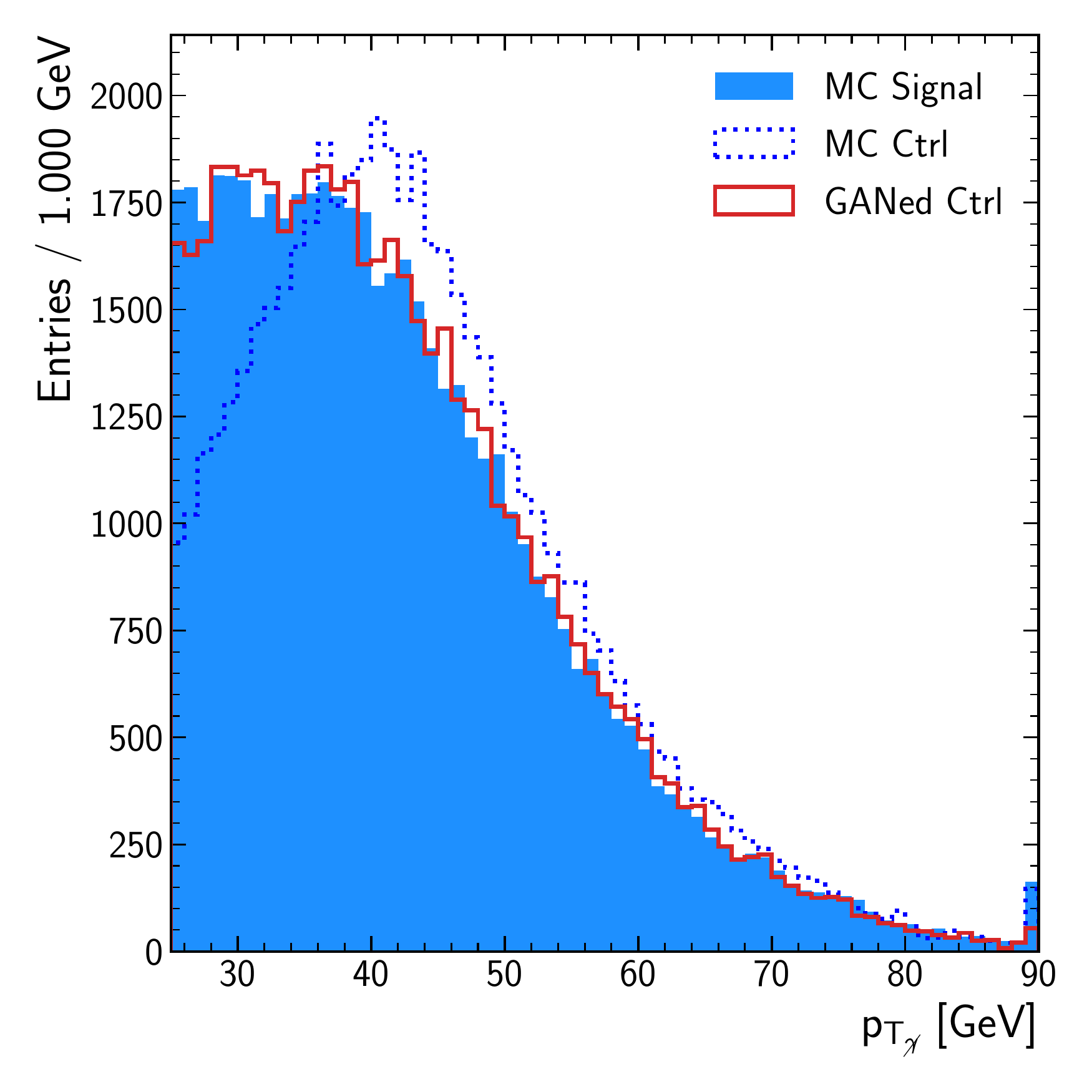}
}

\caption{Distribution of the misidentified photon \pt{} in the MC signal region (MC Signal), MC control region (MC Ctrl) and a misidentified photon generated by the GAN using observables of the events in the the control region (GANed Ctrl).}
\label{fig:1Dpt}       
\end{figure}

\begin{figure*}[ht]

\resizebox{0.95\textwidth}{!}{%
  \includegraphics{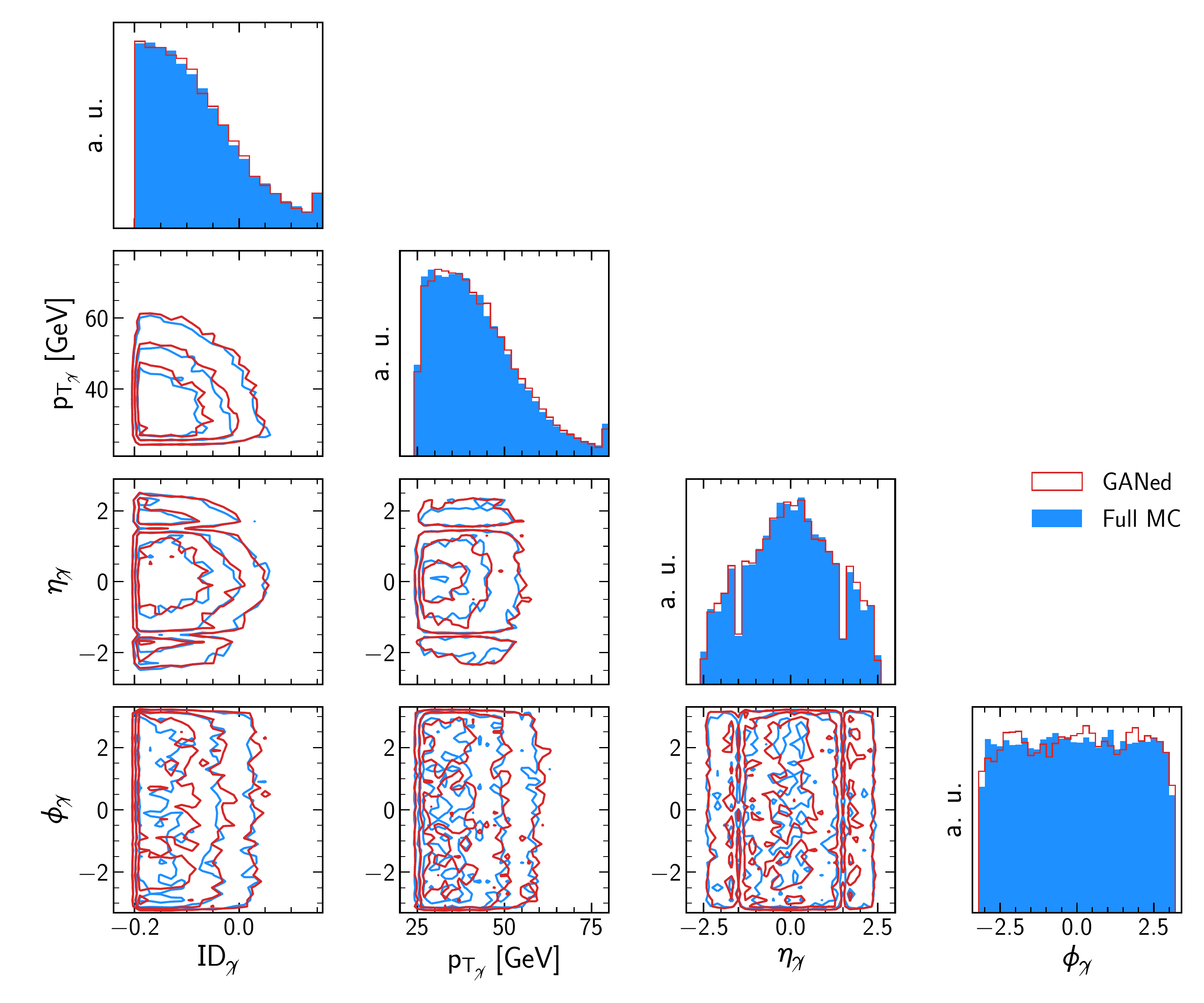}
}

\caption{Diagonal: Comparison of the misidentified photon-observable distributions from the MC \SR (named Full MC) and from the application of the GAN to the MC \CR (named GANed). Off diagonal: Contour lines containing 25~\%, 50~\% and 75~\% of the events and showing correlations between pairs of observables.}
\label{fig:2Dgrid}       
\end{figure*}

As described in the previous sections, we select a \SR with two photons passing stringent photon \id{} criteria. 
A \CR composed of events with one photon candidate passing the \id{} criteria and another one not passing them is formed. As the second object fails to pass the photon selection criteria, it is likely to originate in a jet. This latter object is replaced with a misidentified photon~\misgam generated by the GAN model, thus with \SR properties. The striking transformation capability of this technique is demonstrated in Fig.~\ref{fig:1Dpt}. The \pt{} distribution of the GANed~\misgam matches the distribution of the same observable in the SR, while the MC-simulated misidentified object from the \CR has different characteristics. 
Figure~\ref{fig:2Dgrid} shows an excellent agreement of the GANed object observable distributions (named GANed in Fig.~\ref{fig:2Dgrid}) compared to the ones from actual misidentified photons from the signal region (named Full MC in Fig.~\ref{fig:2Dgrid}). Furthermore, the fact that the isopleths match between the two distributions indicates that the generator also reproduces the correlations between GANed and original Full MC observables.

\begin{figure*}[t]

\resizebox{0.49\textwidth}{!}{%
  \includegraphics{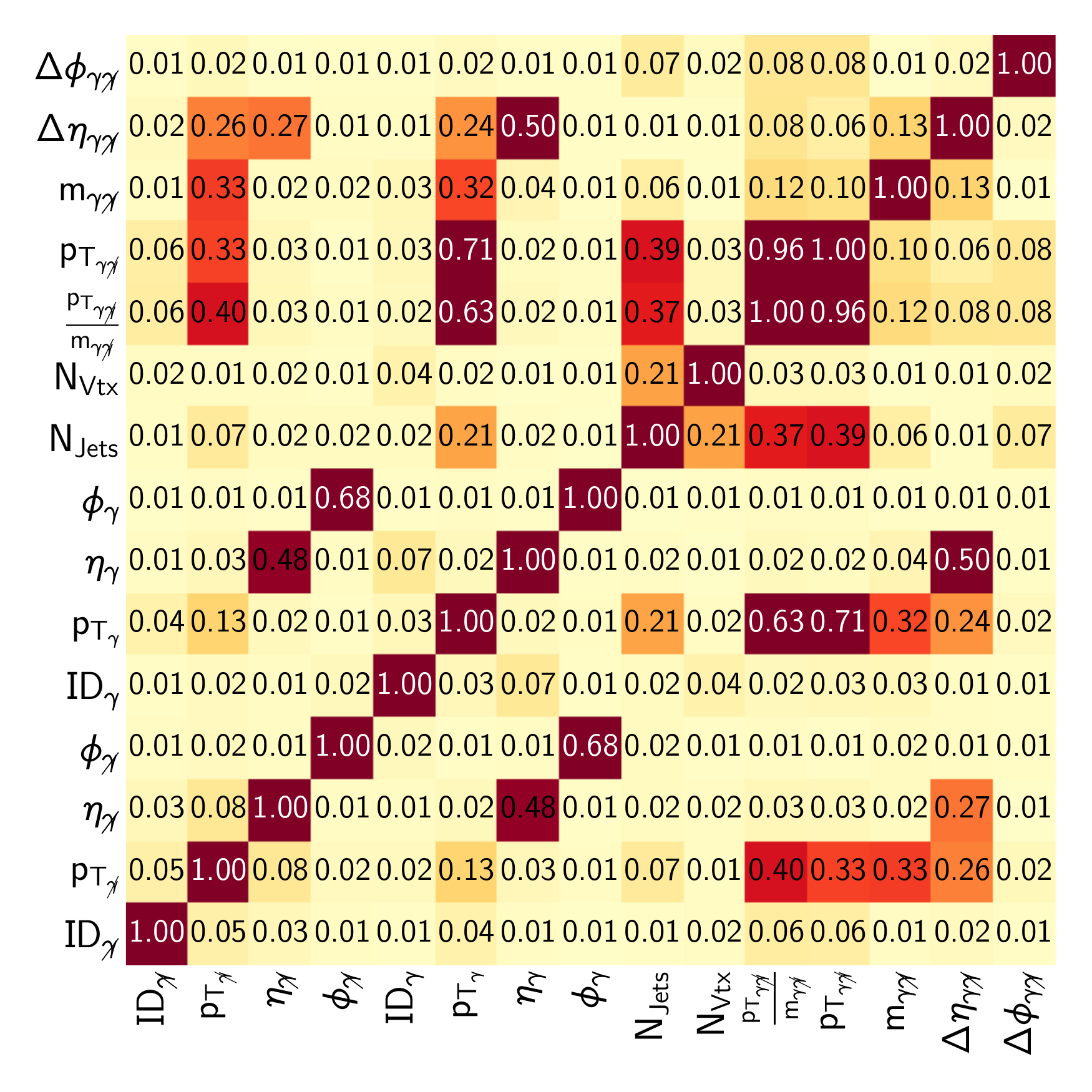}
}
\resizebox{0.49\textwidth}{!}{%
  \includegraphics{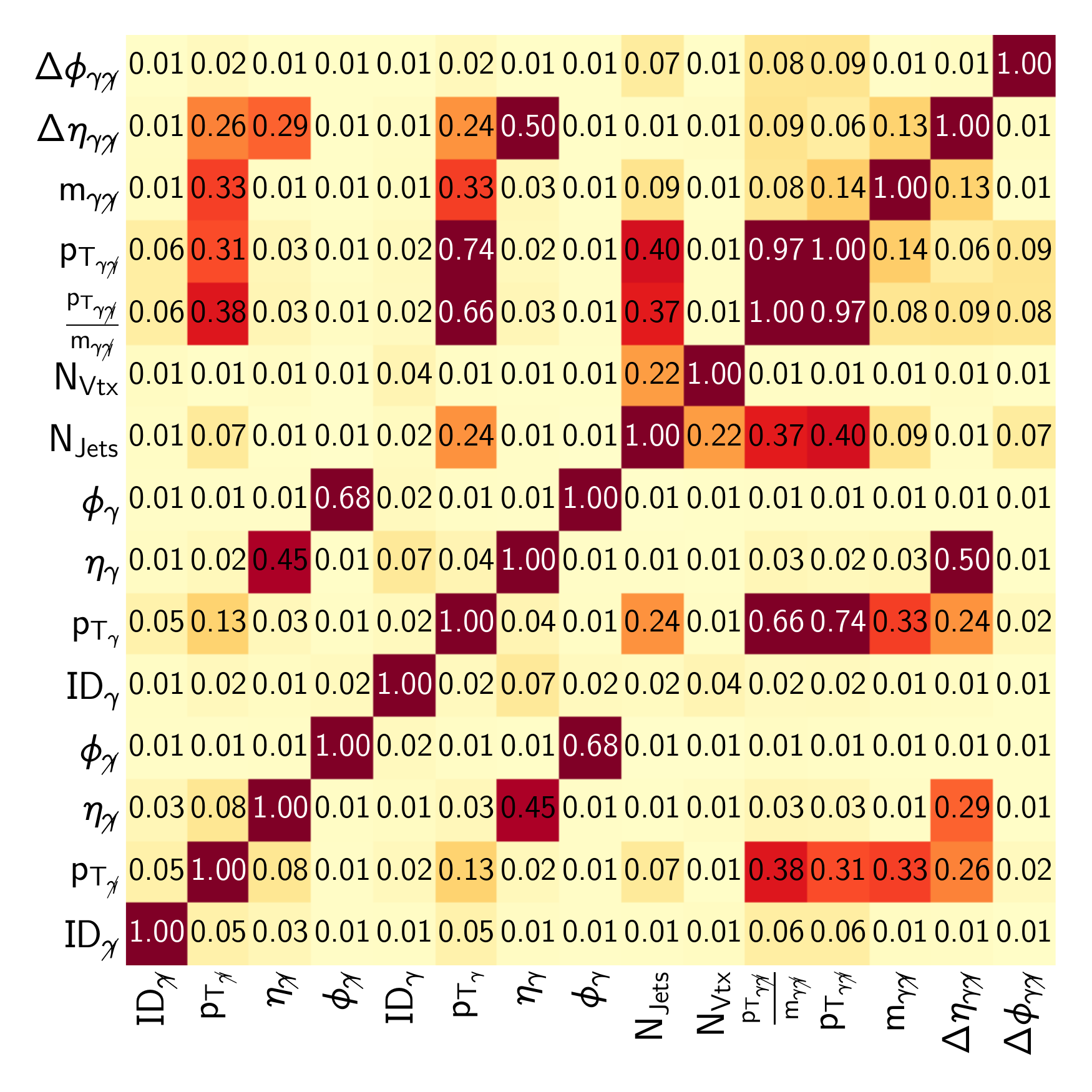}
}
\caption{Correlation matrices computed using the distance correlation on the MC \SR (left) and on sample produced from the GAN application to the MC \CR (right).}
\label{fig:DistanceCorr}       
\end{figure*}

In order to assess the performance of the GAN in terms of reproduction of the correlations between observables, we use the distance correlation as defined in Ref.~\cite{distance_correlation}. This metric allows us to quantify both the non-linear and linear correlations between the observables of the event.
We measure the correlations between the misidentified photon properties and other event observables (prompt photon properties, $N_\jets$, $N_{\mathrm{vtx}}$...) for both the \SR and the \CR with a GANed misidentifed object. These correlation matrices are shown in Fig.~\ref{fig:DistanceCorr}.  
We use a $\chi^2$ to evaluate the difference between the 2 matrices defined as:
$$\chi \equiv \frac{1}{N_{\chi^2}}\sum_{i<j} \sqrt{\chi^2_{ij}} \quad \mathrm{with} \quad \chi^2_{ij} \equiv\frac{\left(d_{ij}^{\CR} - d_{ij}^{\SR}\right)^2}{\sigma_{ij}^2}\ \mathrm{,}$$
where $(i,j)$ is a pair of observables, $d_{ij}$ the corresponding distance correlation, $\sigma_{ij}$ the corresponding statistical fluctuations due to the finite size of the sample, and $N_{\chi^2}$ the number of couples with $i<j$. Correlations are well reproduced and compatible with originating from statistical fluctuations as $\chi \approx 1.1$. While most of the individual $\sqrt{\chi^2_{ij}}$ are below 3, a few exceptions might be noted: the correlations between \pt{\misgam} and $N_\jets$ are at the level of $\sqrt{\chi^2_{\pt{\gam\misgam}\,N_{\jets}}}\approx 5$, denoting some degree of imperfection in the GAN generation.

\section{Conclusion}
\label{sec:conclusion}
In this paper, we presented a new data driven technique to create background samples for HEP background processes with one misidentified object. A \CR is defined by requiring an object to fail \id{} criteria. This object is replaced by a generated misidentified object to simulate an event in the \SR. The technique is based on conditional generative adversarial networks (GANs), known to be difficult to train. To assess the performance of the generator, we developed a figure of merit based on a log-likelihood. Due to random fluctuations intrinsic to the latent space used in GANs, we introduced a multiple-sampling method to obtain more consistent results in the model-performance evaluation.\\

We demonstrated the application and the performance of the technique for the \gj background in the context of the \hgg analysis at the LHC. We have shown that the conditional GAN based technique not only produces object observables that have excellent agreement with the signal-like object observables, but also non-linear correlations of these observables within themselves and with the properties of the rest of the event. Therefore, the samples generated by this technique can be used to improve the description of this background in the \hgg analysis. This is especially true in \SR with specific constraints where the MC simulation might be suboptimal, due to its lack of statistics and/or its inaccurate description of the \gj background.

\acknowledgement{\textbf{Acknowledgements} We gratefully acknowledge the CMS collaboration for the access to the CMS simulation datasets, the CEA (France) for financial and computing support, the Université Paris-Saclay (France) for financial support. We also thank Dr. James Rich and Dr. Nathalie Besson for their careful reading of this manuscript.}

\acknowledgement{\textbf{Data Availability Statement} The datasets generated during and/or analysed during the current study are available in Ref.~\cite{hgg_opendata}.}

\acknowledgement{\textbf{Declarations}}

\acknowledgement{\textbf{Conflict of interest} The authors declare that they have no conflict of interest.}

\acknowledgement{\textbf{Open Access} This article is licensed under a Creative Commons Attribution 4.0 International License, which permits use, sharing, adaptation, distribution and reproduction in any medium or format, as long as you give appropriate credit to the original author(s) and the source, provide a link to the Creative Commons licence, and indicate if changes were made. The images or other third party material in this article are included in the article’s Creative Commons licence, unless indicated otherwise in a credit line to the material. If material is not included in the article’s Creative Commons licence and your intended use is not permitted by statutory regulation or exceeds the permitted use, you will need to obtain permission directly from the copy- right holder. To view a copy of this licence, visit \url{http://creativecomm ons.org/licenses/by/4.0/}.\\
Funded by SCOAP$^3$. SCOAP$^3$ supports the goals of the International Year of Basic Sciences for Sustainable Development.}
%
%
\bibliographystyle{ieeetr}
\bibliography{reference}
%
%
%

\end{document}